\documentclass[sigconf]{acmart}
\usepackage{makecell}
\usepackage{soul}
\usepackage{multirow}

\begin{document}

\title{Mellum: Production-Grade in-IDE Contextual Code Completion with Multi-File Project Understanding}

\author{Nikita Pavlichenko}
\affiliation{%
  \institution{JetBrains}
  \city{Berlin}
  \country{Germany}}
\email{nikita.pavlichenko@jetbrains.com}

\author{Iurii Nazarov}
\affiliation{%
  \institution{JetBrains}
  \city{Munich}
  \country{Germany}}
\email{iurii.nazarov@jetbrains.com}

\author{Ivan Dolgov}
\affiliation{%
  \institution{JetBrains}
  \city{Berlin}
  \country{Germany}}
\email{ivan.dolgov@jetbrains.com}

\author{Ekaterina Garanina}
\affiliation{%
  \institution{JetBrains}
  \city{Yerevan}
  \country{Armenia}}
\email{ekaterina.garanina@jetbrains.com}

\author{Dmitry Ustalov}
\affiliation{%
  \institution{JetBrains}
  \city{Belgrade}
  \country{Serbia}}
\orcid{0000-0002-9979-2188}
\email{dmitry.ustalov@jetbrains.com}

\author{Ivan Bondyrev}
\affiliation{%
  \institution{JetBrains}
  \city{Amsterdam}
  \country{The Netherlands}}
\email{ivan.bondyrev@jetbrains.com}

\author{Kseniia Lysaniuk}
\affiliation{%
  \institution{JetBrains}
  \city{Bremen}
  \country{Germany}}
\email{kseniia.lysaniuk@jetbrains.com}

\author{Evgeniia Vu}
\affiliation{%
  \institution{JetBrains}
  \city{Berlin}
  \country{Germany}}
\email{evgeniia.vu@jetbrains.com}

\author{Kirill Chekmenev}
\affiliation{%
  \institution{JetBrains}
  \city{Amsterdam}
  \country{The Netherlands}}
\email{kirill.chekmenev@jetbrains.com}

\author{Joseph Shtok}
\affiliation{%
  \institution{JetBrains}
  \city{Prague}
  \country{Czech Republic}}
\email{joseph.shtok@jetbrains.com}

\author{Yaroslav Golubev}
\affiliation{%
  \institution{JetBrains Research}
  \city{Belgrade}
  \country{Serbia}}
\email{yaroslav.golubev@jetbrains.com}

\author{Anton Semenkin}
\affiliation{%
  \institution{JetBrains}
  \city{Belgrade}
  \country{Serbia}}
\email{anton.semenkin@jetbrains.com}

\author{Uladzislau Sazanovich}
\affiliation{%
  \institution{JetBrains}
  \city{Munich}
  \country{Germany}}
\email{uladzislau.sazanovich@jetbrains.com}

\renewcommand{\shortauthors}{JetBrains}

\begin{abstract}
We present the Mellum models family, open-weight code completion models designed for interactive use in JetBrains IDEs. 
Mellums have 4B parameters, adopt a Llama-style architecture, and are pre-trained on ~4T tokens of permissively licensed, multi-language code. 
Our studies show that (i)~careful data curation and staged training significantly improve the model's quality, (ii) editor-critical capabilities such as context packing are necessary for high-quality suggestions, and (iii) a compact, task-focused model can meet the cost and latency constraints of interactive completion.

In the paper, we describe an end-to-end industrial pipeline for producing contextualized in-editor completion: disciplined data governance, multi-stage training that includes fill-in-the-middle and project context via supervised fine-tuning, and alignment via direct preference optimization using feedback from real-world scenarios. 
Our quality evaluations include both large-scale offline benchmarks and online telemetry from production deployments in JetBrains IDEs. 
Mellums are released under the Apache-2.0 license on HuggingFace, with a public model card providing a reproducible reference for practitioners. 
Our experience offers a pragmatic blueprint for taking a focused, open model from a research prototype to at scale production for hundreds of thousands of users.
\end{abstract}

\ccsdesc[300]{Computing methodologies~Model development and analysis}
\ccsdesc[300]{Computing methodologies~Neural networks}
\ccsdesc[100]{Computing methodologies~Natural language processing}

\keywords{code completion, IDEs, LLMs, context composing}

\maketitle

\section{Introduction}

Code completion remains one of the most impactful features of modern IDEs because it reduces boilerplate typing, accelerates navigation of unfamiliar APIs, and keeps developers in flow~\cite{meta-completion, completion-cost-benefit}.
In industrial IDEs, completion as a feature spans from purely lexical suggestions to sophisticated engines that incorporate static analysis and semantic context~\cite{completion-abbreviated, vs-usage}. 
Historically, however, most built-in IDE engines have focused on \emph{single-token} suggestions, while \emph{multi-token}, inline suggestions (``gray text'') have only recently become mainstream. 

At JetBrains, we aim to increase the productivity of developers, providing them with tooling that simplifies and speeds up software development.
Prior work at JetBrains described how we brought multi-token locally-run \emph{Full Line Code Completion} (FLCC) to IDEs~\cite{full-line-pipeline}, including its evaluation and the constraints of running locally on end-user machines.
Recently, the broader AI-for-code landscape has shifted toward large, general-purpose LLMs and a wave of ``open-weight'' code models~\cite{codellama,codestral,qwen25-coder,deepseek-coder-v2}. 
Such approaches allow for generating larger code snippets, increasing the quality of solutions for code completion.
While attractive on paper, off-the-shelf architectures have proven ill-suited for production-grade code completion in IDEs for three practical reasons. 

\begin{itemize}
    \item \textbf{Licensing.} Many widely used open models are shipped under use-restricted licenses (\textit{e.g.}, Llama's community license and additional commercial terms~\cite{llama-license}; BigCode's OpenRAIL-M restrictions~\cite{bigcode-license}; and non-production licenses such as Codestral's~\cite{mistral-license}), complicating redistribution and enterprise adoption. 
    
    \item \textbf{Performance/UX.} In addition to inconsistent output format~\cite{jetbrains-mellum-post}, chat-oriented LLMs typically incur high serving costs and latency.
    Also, they often lack editor-critical behavior, for instance, accounting for the code below the caret and partial tokens robustness~\cite{full-line-pipeline}, which are necessary for on-the-fly in-editor completion~\cite{codellama, incoder-paper}.
    
    \item \textbf{Model governance.} For IDE vendors, the lack of transparency about training corpora and irregular model updates can increase the risk of concept drift. 
\end{itemize}

To address these issues, at JetBrains, we developed \textbf{Mellum models family} -- purpose-built (``focal'') models for multi-line code completion that we trained from scratch on permissively licensed public code, and later released openly. 
Mellums are designed specifically for in-editor completion rather than general dialog, which implies their industrial constraints:
\begin{itemize}
    \item Retain \textbf{low latency} to power up real-time code completion suggestions in the IDE.
    \item Keep a \textbf{reasonable model size}, so that a Mellum model together with processed data fit into cost-efficient GPUs -- typically used hardware for LLMs.
    \item Leverage \textbf{widely adopted architecture} to use optimized training and inference frameworks.
\end{itemize}
Having satisfied the constraints above, we are able to power cloud completion for hundreds of thousands of users, and this organically complements our existing local completion stack in JetBrains products.

Most production IDE assistants today are delivered as closed services with undisclosed or restricted models. 
GitHub Copilot, Cursor, and Windsurf exemplify this: they either route requests to a rotating set of third-party frontier models (\textit{e.g.}, OpenAI, Anthropic, Google) or operate over specialized in-editor autocompletion (``Tab'' models) like Cursor's Tab~\cite{cursor-tab} or Windsurf's SWE-1 family~\cite{windsurf-swe}.
These models are exposed via product APIs and UI, not as open weights suitable for redistribution or self-hosting. 
For practitioners, Mellums change the situation: instead of choosing between high-quality but closed tools and hobby-scale open models, Mellums offer an industrial, reproducible reference that can be fine-tuned and integrated under standard OSS compliance workflows.

This paper focuses on the \textit{industrial} problem of training and validating a \textit{contextualized} code completion model and shipping it to a large IDE ecosystem under real-world constraints. 
Overall, our contributions are the following:

\begin{itemize}
    \item Open-sourcing Mellum models as 4B-parameter models on HuggingFace~\cite{jetbrains-mellum-post, jetbrains-mellum-hf}, licensed under Apache 2.0, with public model cards and baseline benchmarks.
    Published instances include the base Mellum model, Mellums fine-tuned for code completion on different language subsets, and additional better-quality installations that went through direct preference optimization stage.
    
    \item An end-to-end training and data pipeline for \textit{contextualized} multi-line code completion: large-scale pre-training on permissively licensed code, structured fill-in-the-middle (FIM) and multi-file context construction using an internal context engine, and alignment with direct preference optimization to suppress unhelpful generations.
    
    \item Evidence that careful data handling and multi-stage training measurably improve completion quality: we report \emph{offline} results at scale for open-sourced Mellums and \textit{online} metrics from production IDE usage of similar Mellum deployment across hundreds of thousands of developers.

    \item A pragmatic framing of the code completion problem in industry, describing production-grade constraints, and explaining why generic open-weight architectures are often unsuitable for production needs, bridging the gap between research and industry effectively.
\end{itemize}

In sum, Mellum shows that a compact, task-specialized model coupled with disciplined data curation and evaluation can deliver practical, production-quality code completion. 
Beyond describing model training, data, and evaluation pipelines, we aim to bridge industry and academia by detailing the decisions, constraints, and feedback loops that take a multi-stage model from research prototype to production in a large IDE ecosystem.

\section{Related Work}

\subsection{Industrial Context}
Industrial adoption of multi-token inline code completion has accelerated in the past years. 
Tools like GitHub Copilot, Cursor, and Windsurf became extremely popular among programmers, transforming software development pipelines and industry as a whole.

Within IDEs, JetBrains' \emph{Full Line Code Completion} (FLCC) documents an end-to-end stack for model-based, locally served, multi-token completions that are latency-aware and constrained to produce syntactically valid insertions~\cite{full-line-pipeline}. 
The paper details product constraints (cold-start memory, token budgets, UI/UX acceptance thresholds) and reports both offline evaluation results and large-scale online telemetry, emphasizing that completion engines must be robust to partial tokens, formatting boundaries, and frequent context shifts typical of editing workflows. 
Industrial restrictions motivate careful prompt scaffolding, short context window~\cite{full-line-context}, sophisticated caching strategy, and compact model size. 

\subsection{Code LLM Families}
From the perspective of model-based solutions for the code completion task, several major model families have recently emerged.
We provide a brief overview of the models to set a baseline for further comparisons.

\paragraph{Code Llama}
Code Llama extends Llama-2 with code-specialized pre-training, long-context conditioning, and built-in infilling capabilities (\textit{i.e.}, fill-in-the-middle, or FIM)~\cite{codellama}. 
Its variants (base, Python, instruct) highlight a general industrial pattern: re-purposing a general LLM into a code-focused family ranging from 7 to 70 billion parameters. 
Code Llama's release set a strong baseline for open-weight code models used in IDE plugins and services. 
License terms (Meta community license~\cite{llama-license}), however, complicate redistribution -- one of the frictions motivating domain-specific, permissively licensed alternatives.
Additionally, the training dataset and the code for training are not public, which implies its own limitations for businesses.
Finally, models come in sizes not suitable for performance and cost-efficient production deployments for real-world usage.

\paragraph{Qwen2.5-Coder}
Qwen2.5-Coder~\cite{qwen25-coder} presents a family of open-weight code-specialized models (0.5B--32B) pre-trained on $\approx$~5.5T tokens with extensive data cleaning, synthetic data generation, and balanced mixing. 
The technical report and model cards emphasize multi-language coverage, code reasoning and repair, and (for larger variants) long-context support (up to 128K in the 32B release), positioning the series as a general-purpose code backbone adaptable to IDE features and agents. 
From an IDE-completion standpoint, the strengths are scale, breadth, and long-context capacity.
Practical challenges include steering away from chat-like formats and enforcing strict insertion formatting without wrappers.

\paragraph{DeepSeek-Coder}
DeepSeek-Coder~\cite{deepseek-coder} demonstrated an even stronger example of a coding models family. 
Authors trained several models ranging from 1.3 to 33B parameters specifically for code-related tasks, including code completion with fill-in-the-middle support.
Their work showed that a task-specific model significantly outperforms general-purpose open- and closed-source LLMs, while preserving a more compact size.
Later on, they continued the work with DeepSeek-Coder-V2~\cite{deepseek-coder-v2}, leveraging a Mixture-of-Experts architecture, and scaling pre-training by $\approx$~6T additional tokens with an up to 128K context window.
DeepSeek's long-context throughput and breadth of language support are attractive for repository- and project-level tasks that require scanning many files and APIs. 

\paragraph{Codestral}
Mistral's Codestral (22B)~\cite{codestral} is a code-oriented open-weight model with explicit FIM support, a 32K context window, and public results on Python and repo-level tasks. 
The release places particular emphasis on repository-scale evaluation (\textit{e.g.}, RepoBench~\cite{repobench-paper}) and on developer integrations via a dedicated endpoint and tool ecosystem. 
From an adoption perspective, its Non-Production License makes the model easy to test but gated for commercial redistribution, again highlighting a tension between quality and licensing that integrators must navigate.

\paragraph{Other models}
StarCoder2~\cite{starcoder2-the-stack2-paper} advances open code LLMs with a rigorously governed dataset (The Stack v2), opt-out/PII removal processes, and long-context training; it provides strong baselines for permissive, open-science releases. 
Earlier, InCoder~\cite{incoder-paper} introduced zero-shot code infilling with bidirectional conditioning, catalyzing FIM as a first-class pre-training/inference mode for code LLMs. 

These works shape the methodological baseline Mellum builds on: task-specific focus, infilling-centric training/inference, and strict data governance. 

\subsection{Benchmarks for Code Completion and Repository-Level Modeling}

Many works in academia focus on evaluation benchmarks for code-related tasks and code completion in particular, rather than on model training.
Here we provide a brief overview of several benchmarks that focus on code, models' infilling capabilities, and project context awareness.

\paragraph{Coding capabilities evaluation.}
HumanEval~\cite{humaneval-codex-paper} remains the canonical functional-correctness benchmark for Python code synthesis. 
Its simplicity enabled rapid iteration and clear \texttt{pass@k} metrics, but it also led to saturation, sensitivity to contamination, and underspecified tests.
Recent extensions -- including HumanEval+ and MBPP+ in EvalPlus~\cite{evalplus-paper} -- expand test coverage to curb false positives and re-rank models more reliably. LiveCodeBench~\cite{livecodebench-paper} further addresses contamination by continuously sourcing fresh tasks across competitive platforms. 
These developments caution against relying on small static test suites when drawing conclusions about IDE completion quality.

\paragraph{FIM-specific evaluation.}
SAFIM~\cite{safim-paper} formalizes syntax-aware FIM evaluation across 17,720 examples and three sub-tasks (block, control-flow expressions, API call completion), sourced post-2022 to mitigate data contamination.
The benchmark design (AST-aware post-processing, prompt templates, syntax-normalized scoring) isolates infilling proficiency from left-to-right generation. 
Results reported by the authors suggest that FIM-pre-training boosts both infilling and left-to-right completion.
Also, they demonstrate that pre-training strategy and data quality can outweigh raw parameter counts, which aligns with industrial observations that IDE-grade completion benefits from targeted pre-training and alignment.

Additionally, HumanEval-Infilling re-purposes the original HumanEval~\cite{humaneval-codex-paper} by masking a contiguous span from the ground truth solution and requiring the model to generate the missing code given the prefix and suffix. 
The public harness executes the completed functions against the unit tests and reports \texttt{pass@k}, expecting code-only outputs -- useful for checking whether models avoid chat-style wrappers. 
The resulting test sets are substantially larger than HumanEval’s 164 problems, which lowers variance and makes infilling gains easier to detect. 
The FIM study in this paper also shows that character-level span selection better exposes token-boundary artifacts at insertion points, a frequent source of IDE glitches. 

\paragraph{Repository-level evaluation.}
To evaluate cross-file reasoning and retrieval, RepoEval~\cite{repocoder-repoeval-paper} packages real repositories with unit tests and three levels of granularity -- line, API invocation, and function body -- supporting executable assessment and time-stamped snapshots to reduce leakage. 
The companion RepoCoder framework demonstrates that iterative retrieval-generation can beat in-file baselines.
Additionally, Long Code Arena~\cite{long-code-arena} proposes a packed evaluation suite  with repository-level contexts for multiple coding tasks, including repository-level code completion.

\subsection{Takeaways for IDE-Grade Completion}
Across industrial reports and research models, three topics recur. 
First, IDE-grade completion hinges on \textit{editor-critical} behaviors -- FIM, strict insertion formatting, and mitigation of token-boundary artifacts -- not always prioritized by chat-oriented LLMs. 
Second, \textit{context packing} and retrieval from multi-file projects are decisive for quality at realistic latency; repository-level benchmarks (RepoEval/RepoBench) and FIM-specific suites (SAFIM) better predict in-editor quality than small standalone synthesis sets. 
Third, \textit{licensing and data governance} shape deployability as much as raw accuracy, influencing whether models can be redistributed, fine-tuned, and audited in enterprise IDE ecosystems. 

Mellum's design choices -- permissive data governance, FIM-centric multi-stage training, and alignment for not chat-like inline suggestions -- are thus complementary to the current wave of general code LLMs, focusing specifically on the UX and operational constraints of interactive completion. 
\section{Training Pipeline}
\label{sec:training}

In this section, we cover the entire pipeline for Mellum: architecture and dataset choices, model specifications, and steps that allow for fill-in-the-middle and model's awareness of project context.
We include a detailed description of base model training for general code understanding, supervised fine-tuning for the code completion task, and direct-preference optimization for better alignment with real-world scenarios.

To start with, our goal was to train a model suitable for production-grade multi-line code completion, which implies the following constraints, already mentioned above:
\begin{itemize}
    \item \textbf{Low latency} for real-time code completion suggestions in the IDE, with 90\% of requests served within 500 ms.
    \item \textbf{Reasonable model size}, so that Mellum together with data batches fit cost efficient GPUs -- typically used hardware for LLMs. 
    Our target deployment hardware needed to provide at least 80 GB of VRAM to host the model and the processed data.
    \item \textbf{Widely adopted architecture} to use optimized training and inference frameworks.
\end{itemize}
To satisfy these constraints, we used a scaled-down version of the Llama architecture with 4 billion parameters, so it is fully compatible with high-load inference on Nvidia H100 GPUs. 
The architecture is well-adopted in academia and industry, so there are available solutions for fast inference like the vLLM library~\cite{vllm}.

We created a custom tokenizer (49,152 tokens) based on the dataset that includes both code and natural text data, with a bigger share of code examples than usual. 
We kept the context size of 8,192 both for pre-training and fine-tuning steps.
The model itself is a LLaMA 2 model with the following parameters: 24 attention and KV heads, 30 layers, hidden size of 3,072, and MLP hidden size of 8,256.

\subsection{Data}
Since our goal was to create a code completion model, our raw data included datasets with code-specific data: The Stack~\cite{the-stack}, The Stack v2~\cite{the-stack-v2}, StarCoder data~\cite{starcoder-data}, RosettaCode~\cite{rosetta-code}, CommitPack~\cite{commit-pack}, and CodeNet~\cite{code-net}. 
For general knowledge and basic natural language understanding, we also included the Wikipedia dataset~\cite{wikipedia-data} to allow for a better completion of comments and string literals. 
Unfortunately, these datasets are not fully up to date, so we addressed this by collecting additional data with fresh open-source code. 

After collecting raw data, we applied file-level filtering by permissive licenses and cleaned it of personal identifiable information (PII) using the Starcoder-PII model~\cite{starcoder-paper}.

\subsection{Pre-training}
The goal of the pre-training stage is to introduce the model to a wide variety of languages, make it learn the syntax, patterns, programming concepts, as well as general knowledge.

We sampled our combined dataset multiple times to reach approximately $4$ trillion tokens.
Data examples were split into chunks of matching size. 
Then, for half of the files in each chunk, we apply the fill-in-the middle transformation.
FIM examples are split into three random-sized parts -- prefix (P), middle (M) and suffix (S) -- that are rearranged into S-P-M order for each particular sample.
The order represents the left-to-right prediction of the middle part given the suffix and the prefix as context.
The resulting mix of raw and FIM data represents a complete dataset for the pre-training step of the model.

The pre-training was conducted on a cluster of 32 nodes with eight H100 GPUs each, and took about 14 days to complete. 
As a result of the pre-training stage, we acquired the \textbf{Mellum-4B-base} model, which is open-sourced on HuggingFace~\cite{jetbrains-mellum-hf}.

\subsection{Supervised Fine-tuning}
After the base model is trained, we transition to a supervised fine-tuning (SFT) stage.
The main objective for the SFT is to tune the model to a specific downstream task, which in our case is code completion. 
For that, we used the same raw datasets but pre-processed them differently: we created better fill-in-the-middle examples and added repository-level contextual information. 
This allowed the model to shape generations' scope to real-world cases and leverage project-level context.

\subsubsection{Better FIM}
\begin{figure*}[] 
    \centering
    \includegraphics[width=\textwidth]{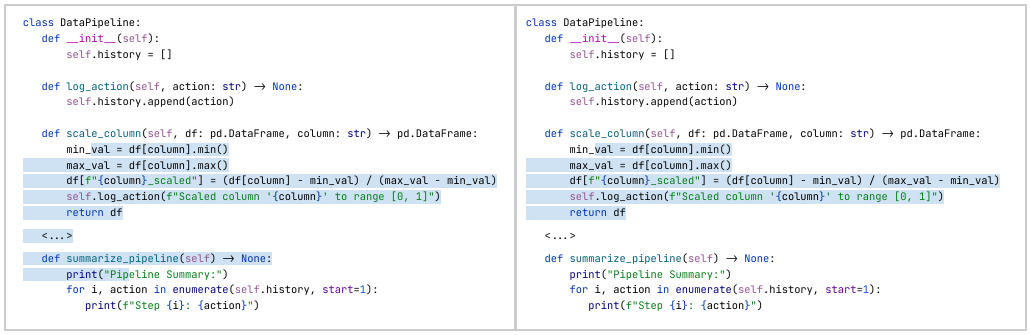}
    \Description{Examples of fill-in-the-middle splitting: (left) random and (right) realistic.}
    \caption{Examples of fill-in-the-middle splitting: (left) random and (right) realistic.}
    \label{fig:fim_together}
\end{figure*}

In real-world cases, users do not fill random chunks of code in files -- they rather work on semantically whole parts like implementing a function or a loop body.
So, in contrast to the pre-training step, where prefix, middle, and suffix were randomly sliced, we prepared more relevant samples for code completion as exemplified in Figure~\ref{fig:fim_together}.

Examples were prepared with Code Engine -- an internal lightweight CLI tool for code processing -- selecting proper meaningful \textit{middle} segments like function or loop bodies.
Our approach identifies syntactic boundaries (block boundary, line start, line middle, token middle) and probabilistically decides whether to start or end the middle segment there.
We also limit the middle chunk to 700 characters to avoid excessively long completions.

Such scope-preserving examples keep generation within a clearly delimited region and mitigate the ``jarring effect''~\cite{meta-completion} that occurs when completions cover several semantic chunks. 
On the integration side in JetBrains IDEs, we additionally truncate returned suggestions so that they remain within a single scope rather than spanning through multiple semantic blocks. 
Training the model to stay within the scope thus reduces the generation of unnecessary text the client would discard anyway and improves the perceived stability of inline suggestions.
As we show in Section~\ref{sec:offline-eval}, such an approach helps the model to stop generations when needed, and, thus, substantially improves evaluation scores.

\subsubsection{Project-level Context Collection}
Effective code completion requires relevant contextual information from the user's code base~\cite{repocoder-repoeval-paper}. 
To incorporate project-level context during the SFT stage, we used Code Engine to collect contextual information directly from the plain project directory. 
For the given file, Code Engine searches for the most relevant chunks in the project using different \textit{context collection strategies}. 
We developed and mixed several context collection strategies in our SFT dataset. 
This prevents overfitting to any single strategy and allows flexible experimentation  during inference, including strategies which require user interactions and IDE tooling.

Almost all of our strategies rely on \emph{intersection over union (IoU)} as a similarity metric, which we found to work well for selecting relevant code chunks while being computationally efficient. Calculating this metric requires minimal computational overhead, unlike more expensive approaches such as cosine distance over semantic embeddings.

Formally, for two code snippets $x$ and $y$, we split each snippet into chunks (which may be $n$-grams, tokens, or lines). 
This produces two sets of chunks,
\[
C_x = \{c_i^x\}, \quad C_y = \{c_i^y\}.
\]
The IoU similarity is then defined as
\[
\mathrm{IoU}(x,y) = \frac{|C_x \cap C_y|}{|C_x \cup C_y|}.
\]

We add a subscript to designate the granularity of code chunks. 
For instance, if the snippets are split into byte-pair encodings (BPE), we write $\mathrm{IoU}_{\mathrm{BPE}}(x,y)$. 
Similarly, we may use $\mathrm{IoU}_{\mathrm{line}}(x,y)$ for lines.
Below, we cover some of the most effective strategies we experimented with.

\textbf{IoU Strategy.} 
Our simplest strategy collects files in the same directory as the current file and then selects the closest ones based on the $\mathrm{IoU}_{\mathrm{line}}$ similarity.

\textbf{Path Distance Strategy.} 
To address the limitations of the \textit{IoU strategy} in highly modular projects, we developed the \textit{Path Distance strategy}. 
In such projects, individual modules contain few files, so the most relevant code may reside in related modules rather than within the same directory.

We define \textit{path distance} as the minimum number of directory traversals required to navigate from one file to another in the project hierarchy. 
Our implementation performs a breadth-first traversal of the directory hierarchy, traversing both parent and child directories from the current file. 
This way, the files are collected in order of increasing path distance.

\begin{figure*}[t]
    \centering
    \includegraphics[width=\textwidth]{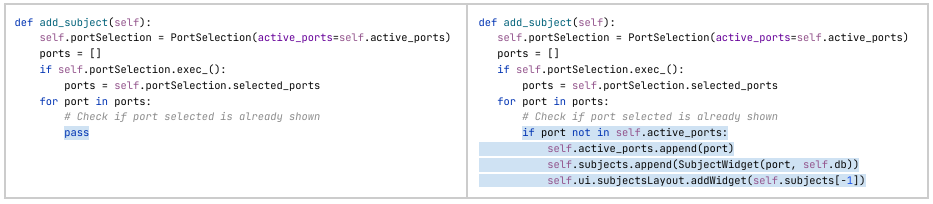}
    \Description{Example of DPO data: a negative sample (left) and a positive sample (right).}
    \caption{Example of DPO data: (left) a negative sample and (right) a positive sample.}
    \label{fig:dpo_together}
\end{figure*}

\textbf{RAG Strategy.}
To address potential code irrelevance from naive collection of files from the repository, we search for the relevant code chunks among the list of files collected via the \textit{Path Distance strategy}.

Overall, the RAG strategy operates as follows:
\begin{enumerate}
    \item Gather candidate files using \textit{Path Distance} strategy.
    \item Split each candidate file into overlapping line chunks using a sliding window.
    \item Extract a context window of the same size as a sliding window around the current cursor position.
    \item Score each chunk using $\mathrm{IoU}_{\mathrm{BPE}}$ similarity against the cursor context.
    \item Select the highest-scoring chunks across all candidate files and include them in the context.
\end{enumerate}

Formally, for a candidate file $f$ split into chunks $\{c_1, c_2, \ldots, c_n\}$ and cursor context $q$, we compute:
\[
\text{score}(c_i) = \mathrm{IoU}_{\mathrm{BPE}}(c_i, q).
\]
The final context consists of the highest-scoring chunks regardless of their source file.

In real-world coding experience, programmers often have several open files that may be semantically connected, \textit{e.g.}, when the function is implemented in one file and used in another.
We additionally include such files during inference to expand the search of the relevant chunks.
Accounting for this during training is challenging, because recent files are only known from user's interaction with the IDE and cannot be derived from static repository datasets.
During the inference, however, we enhance the list of candidate files on step $1$ with \textit{recent files}, which are the files with which the developer has recently interacted in the IDE.

\subsubsection{SFT Results}
As a result of SFT, we acquired several models like \textbf{Mellum-4b-sft-python} (Python-specific data only), \textbf{Mellum-4b-sft-all} (data for all languages), which are open-sourced on HuggingFace~\cite{jetbrains-mellum-hf} as fine-tuned versions of the base model.

\subsection{Direct Preference Optimization}
After the SFT stage, the model adhered closely to the FIM objective and the usage of contexts.
However, it still tended to produce verbose, hard-to-read code and occasionally generated syntactically correct yet unhelpful outputs (e.g., \texttt{NotImplementedError} stubs in empty methods). 
To improve both readability and utility, we constructed a dataset for the direct preference optimization (DPO). 
We started with sampling outputs from the SFT model across multiple model's temperatures to get diverse examples for the same inputs.
Then, we used an LLM-as-a-Judge (LLMaaJ) procedure~\cite{llmaaj} to score different input-output pairs obtained from the same input. 
This allowed us to create a labeled dataset with ``good'' and ``bad'' generations (see Figure~\ref{fig:dpo_together} for examples).
Basically, we refined the SFT corpus to better reflect IDE usage by making FIM prompts and targets more realistic. 
Finally, leveraging the Tree-sitter parser, we segmented large FIM instances into shorter, stylistically preferable snippets. 
During DPO training, the model was trained to directly align with preference labels.
As we show in Section~\ref{sec:offline-eval}, the resulting post-training model produces even more compact, readable code with improved functional relevance.

As a result of this stage, we acquired several models like \textbf{Mellum-4b-dpo-python} (Python-specific data only), \textbf{Mellum-4b-dpo-all} (data for all languages) and others, which are open-sourced on HuggingFace~\cite{jetbrains-mellum-hf} as fine-tuned versions of SFT models.

The overall described pipeline resulted in a family of Mellum models that includes the base model together with SFT and DPO models.
All models have 4 billion parameters and support the context window up to 8K tokens.

\section{Offline Evaluation}
\label{sec:offline-eval}
To validate Mellums' performance on the code completion task, we begin with offline evaluation, which enables comparison with open-source models of similar scale on the pre-defined datasets. 
We examine three key design choices, namely:

\begin{itemize}
    \item Context collection;
    \item Impact of the preference optimization stage;
    \item Language specialization: Python-only versus multi-lingual fine-tuning.
\end{itemize}

It is important to note that we cannot describe all the details of evaluating different models and setups within the scope of this paper.
So, this section covers only the most important choices we made while establishing the evaluation pipeline.
Overall, in this section, we compare Mellums with open-source coding models (Qwen-2.5, Seed-Coder, DeepSeek-Coder, and CodeLlama) in the same weights category. 
Comparisons are made on both proprietary and well-known public benchmarks.

\subsection{JetComplete Benchmark}
To assess the choices made, we primarily rely on a custom proprietary JetComplete benchmark, which mirrors real-world IDE usage patterns. This benchmark employs the same context collection strategies used at inference time, limiting completions to meaningful code segments, and focusing on realistic completion points. 
The benchmark focuses on evaluating FIM code completion, where models predict the missing code segment given surrounding file text as well as project-wide context.

\subsubsection{Dataset}
The benchmark is built from language-specific subsets and has 15 languages in total (we report results on the 8 most popular ones for brevity).
For each language, we use relevant open-source repositories that are excluded from the model's train set.
From each source file, we create FIM examples that split the file into three parts (prefix, middle, and suffix) with the same Code Engine that was used during SFT training described above. 

To additionally ensure the quality of the evaluation examples, we filter out some of them in two stages. 
First, we apply heuristic filters using \texttt{Pygments}~\cite{pygments} tokenization to drop middles dominated by comments, string/numeric literals, or whitespaces. 
Second, we run an LLM-as-a-Judge pass to exclude subtler low-quality cases (\textit{e.g.}, placeholders, anti-patterns, broken code).
These steps ensure we assess performance on realistic, high-signal completions and avoid noisy or irrelevant segments.

After filtering, we sample the remaining examples to ensure diversity and broad coverage. 
Each example carries metadata inherited from its repository and file (\textit{e.g.}, topic, star count, project age, test vs.\ source file), as well as split-level attributes (\textit{e.g.}, boundary type: line start, block start, etc.). 
We stratify across these attributes, so the final dataset is balanced by topic, repository age, repository popularity, file role, and split strategy.

Next, we add project context to every data sample using the context collection strategies described in Section~\ref{sec:training}.
This ensures that our benchmark uses context collection algorithms that work well in IDEs in production, making the evaluation setting closely match real-world usage.

Finally, we eliminate train–test leakage (including repository-, file-, and near-duplicate overlap) to ensure a fair comparison.

The resulting dataset:
\begin{itemize}
    \item Covers 15 languages.
    \item Contains clean, meaningful middle segments.
    \item Offers balanced coverage across topics, repository ages, popularity, and file roles.
    \item Includes per-split context gathered with the production context collection strategies.
\end{itemize}

\begin{table*}[t]
\caption{Overall performance of code completion models. Bold indicates the global best, and also the best among the Mellum models if it is lower.}
\label{tab:eval:overall}
\centering
\begin{tabular}{lrrrrrrrr}\toprule
{\textbf{Model}} &
\textbf{SAFIM} &
\textbf{HumanEval-I} &
\multicolumn{2}{c}{\textbf{RepoBench-C}} &
\multicolumn{3}{c}{\textbf{JetComplete}} \\
\cmidrule(lr){2-2} \cmidrule(lr){3-3} \cmidrule(lr){4-5} \cmidrule(lr){6-8}
& pass@1 & pass@1 & EM & ES & chrF++ & EM & KK score \\
\midrule
Qwen-2.5-Coder-1.5B & 45.79 & 24.00 & 29.42 & 69.37 & 47.93 & 8.46 & 0.29 \\
Qwen-2.5-Coder-3B & 54.71 & 63.38 & 32.84 & 71.34 & 46.36 & 6.86 & 0.27 \\
Qwen-2.5-Coder-7B & 52.82 & 63.66 & 36.43 & 73.95 & 39.77 & 3.16 & 0.19 \\
Seed-Coder-8B-Base & \textbf{60.82} & 71.26 & \textbf{37.46} & \textbf{74.78} & 62.01 & 18.00 & 0.45 \\
DeepSeek-Coder-5.7B & 59.68 & \textbf{75.46} & 35.33 & 73.71 & \textbf{74.46} & 39.59 & 0.62 \\
CodeLlama-7B & 45.00 & 63.38 & 33.55 & 72.00 & 70.71 & 37.91 & 0.54 \\
\cmidrule(r){1-8}
\textbf{Mellum-4b-base} & 49.62 & 45.79 & 27.51 & 67.38 & 63.91 & 17.64 & 0.48 \\
\textbf{Mellum-4b-sft-python} & 47.66 & 56.13 & 27.90 & \textbf{67.76} & 68.72 & 25.15 & 0.56 \\
\textbf{Mellum-4b-sft-all} & \textbf{52.86} & 55.94 & \textbf{28.53} & 65.13 & \textbf{74.22} & 30.36 & 0.63 \\
\textbf{Mellum-4b-dpo-python} & 47.47 & \textbf{56.53} & 25.37 & 64.20 & 70.88 & 37.18 & 0.63 \\
\textbf{Mellum-4b-dpo-all} & 52.17 & 53.76 & 24.43 & 57.70 & 73.64 & \textbf{42.65} & \textbf{0.69} \\
\bottomrule
\end{tabular}
\end{table*}

\subsubsection{Metrics}
We use several metrics to evaluate code completion on the JetComplete benchmark, all chosen to address different aspects of completion quality and real-world usability.
In our report, we include both classic quality assessment metrics and a custom metric that better reflects human perception of the quality of code completion suggestions.

First, we employ \textbf{Exact Match} (EM) as the strictest and most interpretable metric.
The metric is computed by comparing the \textit{generated middle} part with the \textit{ground truth middle} part.
If they fully match, the generation is considered successful.
However, this metric lacks flexibility since code can deviate from the ground truth while remaining correct and readable.

Second, we incorporate the classic \textbf{chrF++} score as it was shown to be the best choice for code completion among the established publicly available text and code similarity metrics~\cite{codegen-quality}.

Finally, we developed a metric specifically tailored for our task -- \textbf{KK score}. 
It is designed to be deterministic, computationally efficient and well correlated with human judgement of good completion suggestion. 
KK score is defined as a proportion of consecutive lines in the completion that have a matching line in the ground truth middle part. 
The matching is performed using normalized Levenshtein distance, where each completion line is considered a match if it is similar enough to at least one line in the ground truth middle section. 
Disregarding line order in the ground truth middle accounts for cases where the order of lines is not fixed semantically, and edit similarity instead of exact match ensures that we allow small changes in the completion.
The score is calculated by counting matching lines sequentially from the beginning until the first mismatch, assuming that a user reads from top to bottom and stops when the quality degrades.

To ensure that our evaluation aligns with real-world use cases, we manually annotated code completion examples with the ``usefulness score'' and assessed existing metrics against these scores.
Compared to several classical text similarity metrics and an LLM-based metric, KK score demonstrated superior performance.
For multi-line completion, KK-score shows Cohen-kappa of 0.61 compared to 0.28 for EM and 0.5 for LLM-as-a-Judge. The gap narrows in single-line completion (EM: 0.49, LLMaaJ: 0.57) but KK maintains its 0.57 score.

\subsubsection{Summary}
The combination of realistic completion scenarios, project-wide context, and human-calibrated evaluation makes JetComplete a reliable proxy for in-IDE performance.

\subsection{Additional Benchmarks}
In addition to proprietary JetComplete, we also evaluate on three public benchmarks to enable comparison with existing work: SAFIM \cite{safim-paper}, HumanEval-Infilling (\textit{HumanEval-I})~\cite{oai-fim-humaneval-i}, and RepoBench-C~\cite{repobench-paper}.
Although these benchmarks may not capture the IDE-specific behaviors that JetComplete evaluates, they provide standardized comparison points.
These benchmarks use different evaluation metrics -- SAFIM and HumanEval-Infilling rely on pass@1 (execution-based correctness), while RepoBench-C uses exact match (EM) and edit similarity (ES). 

\subsection{Experiments}
Here, we cover data-centric design choices made while training Mellum, highlighting the effects of SFT and DPO stages described in Section~\ref{sec:training}.

\subsubsection{SFT impact}
We first examine the impact of adding project-wide context during the SFT stage.
Table~\ref{tab:eval:overall} shows that supervised fine-tuning with context delivers substantial improvements over the base model across all metrics. 
For example, Mellum-4b-sft-all not only demonstrates KK score improvement over the base Mellum model (0.48 vs 0.63) but also surpasses bigger models like Qwen-2.5-Coder-7B, Seed-Coder-8B-Base and DeepSeek-Coder-5.7B.

\begin{figure}[t] 
    \centering
    \includegraphics[width=\columnwidth]{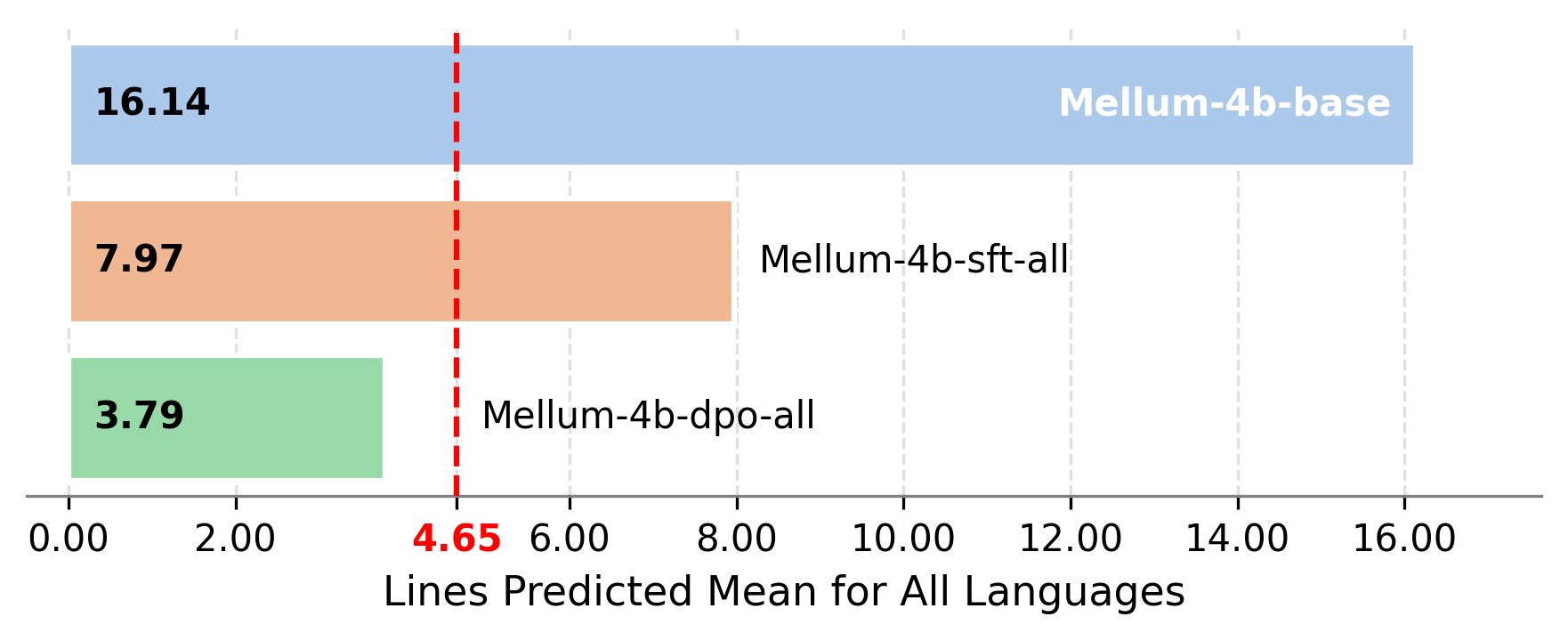}
    \Description{Mean number of predicted code lines for different models, ground truth length in red.}
    \caption{Mean number of predicted code lines for different models, ground truth length in red.}
    \label{fig:mean_lines}
\end{figure}

We also observed that the ability to \textit{stop} is crucial for code completion.
As we see from Qwen models examples in the Table~\ref{tab:eval:overall}, scaling the number of model's parameters often leads to quality degradation.
Our hypothesis is that such an effect is caused by big models generating larger suggestions that may often exceed code block boundaries and, thus, become incorrect.
At the same time, fine-tuned Mellum generates considerably fewer lines than base Mellum (7--8 instead of 16, see Figure~\ref{fig:mean_lines}), which helps to keep the high quality of generated suggestions.

Regarding external open-source benchmarks, SFT yields little to no gain on SAFIM and RepoBench. 
For SAFIM, many tasks include natural-language prompts that provide useful hints. 
Mellums were trained mostly on code, with little text data, so they cannot fully exploit these descriptions.
Comparisons with models trained on mixed code–text corpora are therefore not entirely fair. 
RepoBench, however, focuses on next-line prediction only; it does not test mid-line or mid-token completion, which is common in production for code completion. 
Because evaluation considers only the first generated line, model's stopping behavior is also ignored. 
In addition, RepoBench assumes contexts larger than 8K tokens, while Mellum currently supports up to 8K, which may further limit performance.
As a result, SFT offers limited benefits on these suites and Mellums' scores remain modest yet reasonable.

In contrast to SAFIM and RepoBench, on HumanEval-Infilling, SFT delivers substantial gains. 
Given the benchmark’s strong Python focus, Mellum-4b-sft-python variant performs better than Mellum-4b-sft-all.

\subsubsection{DPO impact}
We next assess alignment to the desired outputs via Direct Preference Optimization (DPO). 
As shown in Table~\ref{tab:eval:overall}, DPO for multi-lingual models provides an additional lift on JetComplete beyond SFT, and a large gain over the base model. 

\begin{table*}[t]
\caption{Performance on JetComplete (KK score).}
\label{tab:eval:JetComplete_kk}
\centering
\begin{tabular}{lrrrrrrrrr}\toprule
\textbf{Model} & \textbf{Python} & \textbf{Java} & \textbf{Kotlin} & \textbf{JS} & \textbf{C++} & \textbf{C\#} & \textbf{Go} & \textbf{PHP} & \textbf{Average} \\ \midrule
Qwen-2.5-Coder-1.5B & 0.30 & 0.29 & 0.35 & 0.34 & 0.28 & 0.28 & 0.25 & 0.27 & 0.29 \\
Qwen-2.5-Coder-3B & 0.25 & 0.28 & 0.32 & 0.31 & 0.26 & 0.27 & 0.23 & 0.28 & 0.27 \\
Qwen-2.5-Coder-7B & 0.17 & 0.21 & 0.23 & 0.20 & 0.20 & 0.17 & 0.16 & 0.18 & 0.19 \\
Seed-Coder-8B-Base & 0.48 & 0.46 & 0.47 & 0.47 & 0.43 & 0.43 & 0.41 & 0.46 & 0.45 \\
DeepSeek-Coder-5.7B & 0.57 & 0.68 & 0.64 & 0.60 & 0.56 & 0.65 & 0.61 & 0.63 & 0.62 \\
CodeLlama-7B & 0.49 & 0.57 & 0.52 & 0.54 & 0.51 & 0.57 & 0.55 & 0.53 & 0.54 \\ \cmidrule(r){1-10}
\textbf{Mellum-4b-base} & 0.48 & 0.49 & 0.48 & 0.46 & 0.46 & 0.52 & 0.45 & 0.46 & 0.48 \\
\textbf{Mellum-4b-sft-python} & 0.61 & 0.58 & 0.56 & 0.56 & 0.51 & 0.58 & 0.52 & 0.54 & 0.56 \\
\textbf{Mellum-4b-sft-all} & 0.61 & 0.67 & 0.68 & 0.62 & 0.57 & 0.67 & 0.60 & 0.63 & 0.63 \\
\textbf{Mellum-4b-dpo-python} & 0.64 & 0.68 & 0.64 & 0.61 & 0.59 & 0.65 & 0.64 & 0.62 & 0.63 \\
\textbf{Mellum-4b-dpo-all} & \textbf{0.68} & \textbf{0.74} & \textbf{0.73} & \textbf{0.67} & \textbf{0.64} & \textbf{0.70} & \textbf{0.69} & \textbf{0.69} & \textbf{0.69} \\
\bottomrule
\end{tabular}
\end{table*}

Importantly, DPO improves stopping behavior: the mean number of generated lines moves closer to the ground truth (Figure~\ref{fig:mean_lines}). 
It is worth noting that this is not merely shorter outputs inflating text distance-based metrics (like ChrF++ or KK score, which is based on Levenshtein Distance), EM also increases for DPO models (Table~\ref{tab:eval:overall}), indicating more exact matches.

On external benchmarks, we do not observe any gains on SAFIM and RepoBench, which is expected given the same reasons described above. 
On HumanEval-Infilling, only Mellum-4b-dpo-python yields a small improvement compared to Mellum-4b-sft-python.

\begin{figure*}[t] 
    \centering
    \includegraphics[width=\textwidth]{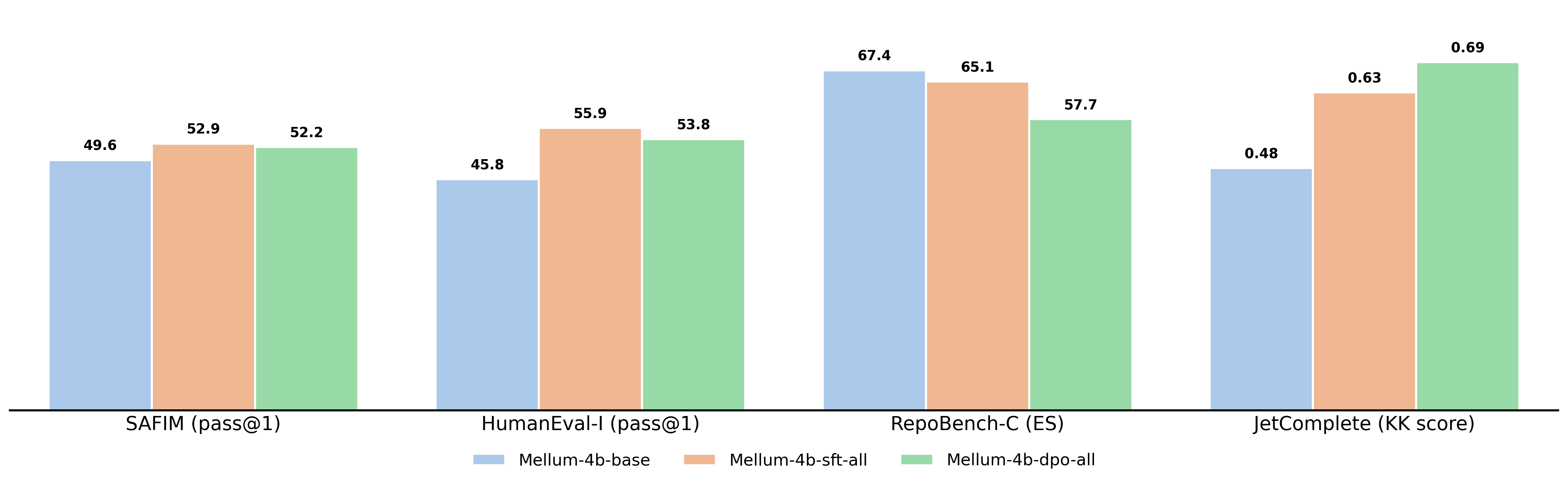}
    \Description{Impact of training stages on the model's quality.}
    \caption{Impact of training stages on the model's quality.}
    \label{fig:sft-dpo}
\end{figure*}

Overall, Figure~\ref{fig:sft-dpo} shows the impact of different training stages on the multi-lingual model quality. 

\subsubsection{Language Specialization}
We evaluate both multi-lingual SFT (many programming languages) and mono-lingual SFT (Python-only) models to provide additional ablation for language-specific training (see Table~\ref{tab:eval:JetComplete_kk}). 
Python-only SFT lifts performance for other languages as well, but multi-lingual SFT yields larger gains while maintaining the same Python performance. 
For DPO, the multi-lingual model (Mellum-4b-dpo-all) shows significant advantages even on Python itself.

\subsection{Key Takeaways}
Overall, SFT and DPO significantly improve Mellum for the code completion setting we target. 
On JetComplete, our task-aligned benchmark, Mellum with SFT and DPO outperforms even larger open-source baselines. 
Results on less aligned public suites are modest but competitive. 
Mellum models also exhibit strong stopping capabilities that are crucial for practical IDE code completion: not only the quality of code completion suggestions increases, but also less unneeded code is generated, leading to more performance- and cost-effective model usage.

\section{Online Evaluation}
In contrast to offline evaluation, online evaluation leverages real users' telemetry instead of pre-defined datasets. 
In our production deployments for cloud code completion in JetBrains IDEs, we run models similar to the open-sourced Mellum family: historically, production Mellums are pre-trained on a smaller amount of data (~3T tokens instead of ~4T tokens). 
Despite the exact models installations being slightly different for production and open-source, they show almost the same offline evaluation results.
So, to further prove the validity of our training approach and demonstrate industrial impact, we briefly cover online metrics in this section.

Following prior work about locally-run code completion for JetBrains IDEs~\cite{full-line-pipeline}, we incorporate Ratio of Completed Code (RoCC) and Acceptance Rate (AR) as the main quality metrics for inline code completion powered by an LLM.
RoCC is defined as a ratio of symbols of code written with code completion (including built-in completion in the IDE) among all the written code in the editor.
AR is the ratio of events where a code completion suggestion was accepted by the user to all events where it was shown.

In addition to quality metrics, we carefully monitor the backend load for the models to understand models' performance.
Our data show that ~400 ms is typically spent for processing a single request, with ~700 ms during peak times throughout the working days.
This aligns well with users' expectations from an in-flow AI-assisted feature.

For the sake of comparison, we share the same up-to-date metrics for locally-run code completion for JetBrains IDEs, which utilizes models ~40 times smaller and runs directly on the end users' devices.
From Table~\ref{tab:eval:online}, we see that cloud code completion with Mellum variant substantially increases RoCC in the IDE editor window and typically increases AR as well.
This highlights large-scale industrial impact of the Mellum models family for professional software developers.

\begin{table}[ht]
\caption{The results of online evaluation for different languages.}
\centering
\begin{tabular}{ccccc}
\toprule
\multirow{2}{*}{\textbf{Language}} & \multicolumn{2}{c}{\textbf{RoCC}} & \multicolumn{2}{c}{\textbf{AR}} \\ \cmidrule(lr){2-3} \cmidrule(lr){4-5}
	& \textbf{local} & \textbf{cloud} & \textbf{local} & \textbf{cloud} \\ 
\midrule
Java   & 0.42 & \textbf{0.49} & \textbf{0.34} &\textbf{ 0.34} \\
Kotlin & 0.32 & \textbf{0.45} & 0.28 & \textbf{0.31} \\
Python & 0.25 & \textbf{0.39} & 0.26 & \textbf{0.34} \\
JS/TS  & 0.28 & \textbf{0.42} & 0.25 & \textbf{0.30} \\
C\#    & 0.37 & \textbf{0.47} & 0.30  & \textbf{0.31} \\
C/C++  & 0.27 & \textbf{0.38} & 0.28 & \textbf{0.32} \\
Go     & 0.37 & \textbf{0.46} & \textbf{0.42} & 0.38 \\
PHP	   & 0.31 & \textbf{0.45} & \textbf{0.32} & 0.31 \\
Rust   & 0.29 & \textbf{0.41} & \textbf{0.33} & \textbf{0.33} \\
\bottomrule
\end{tabular}
\label{tab:eval:online}
\end{table}
\section{Discussion}

\textbf{Production-specific benchmarks.} Our production deployment revealed several key considerations beyond benchmark performance. 
While open-source benchmarks provide useful baselines, selecting models based solely on public leaderboards often fails in production environments. 
Task-specific datasets and validation are essential, along with targeted fine-tuning for the specific use case. 
Our results show that while general language models can be used directly for coding tasks, multi-stage training significantly boosts quality for IDE-specific scenarios, improving KK score from 0.48 to 0.69 in the best version. 
Preference optimization additionally fixes undesired behaviors, like generating \texttt{TODO} comments, empty statements, and over-generation beyond the logical stopping point. 
In the future, user data can be incorporated into the DPO stage to further improve the model behavior and UX. 

\textbf{Are language-specific models worth it?} We experimented with language-specific tuning for Python, but acquired mixed results, observing little to no improvements.
From a deployment perspective, language-specific models become problematic when organizations require self-hosted solutions due to privacy constraints, as costs scale linearly with each additional model.
On top of that, modern developers rarely work with a single programming language, and a polyglot model can use cross-language context that may further improve quality for end users. 
For example, in high-performance Python development, completions can leverage context from parts written in Rust/C/C++. 
Overall in Python evaluations, language-specific model showed no quality gains despite higher training and deployment costs.

\textbf{Navigating constraints}. Similarly, while selecting an architecture with more learnable parameters might have led to better quality on offline benchmarks, sticking to a smaller size allows us to achieve much lower latency, which directly contributes to the perceived quality for our users. 
This is confirmed by strong online results.

Beyond deployment complexity, privacy requirements introduce additional constraints. 
We published quantized versions of our models to enable local deployment for users with the strictest privacy requirements. 
However, most developers lack GPUs and must rely on CPU-only inference, which reaches up to 3 seconds even on the latest hardware, which is an unacceptable latency for interactive code completion.
We believe local deployment will become increasingly viable for latency-critical applications in the future as hardware and software advance.
\section{Conclusion}
We have presented Mellum models family -- focal 4B-parameter models that demonstrate how to bridge a gap between research prototypes and production-ready code completion systems. 
Our studies show that real-world applications require moving beyond public benchmarks optimization and considering data governance, latency, costs, and product requirements like multi-file context awareness. 
Mellums' competitive performance on public benchmarks, as well as superior performance in the task of context-aware code completion, validate this approach -- even bigger models demonstrate lower quality, while requiring more inference-time computations and producing larger outputs.

The success of Mellum serving millions of users in JetBrains IDEs proves that smaller task-specific models can deliver productivity gains while meeting strict product constraints. 
By releasing Mellum under the Apache-2.0 license, we make AI-assisted coding more accessible to organizations that cannot afford large-scale model deployment or have strict privacy concerns limiting usage of API-based LLMs. 
We believe more specialized and efficient models will appear in the future which prioritize practical deployment alongside quality, democratizing AI tools to a broader community.

\bibliographystyle{ACM-Reference-Format}
\bibliography{literature}

\end{document}